\documentclass[11pt]{article}

\usepackage[preprint]{acl}

\usepackage{times}
\usepackage{latexsym}
\usepackage{subcaption}
\usepackage{tabularx}

\usepackage[T1]{fontenc}

\usepackage[utf8]{inputenc}

\usepackage{microtype}

\usepackage{graphicx}

\title{Propaganda is All You Need}

\author{Paul Kronlund-Drouault \\
  \texttt{pkd.unsuspicious.org}
  }

\begin{document}

\maketitle

\begin{abstract}
    As Machine Learning (ML) is still a recent field of study, especially outside the realm of abstract Mathematics and Computer Science, few works have been conducted on the political aspect of large Language Models (LLMs), and more particularly about the alignment process and its political dimension. This process can be as simple as prompt engineering but is also very complex and can affect completely unrelated notions. For example, politically directed alignment has a very strong impact on an LLM's embedding space and the relative position of political notions in such a space. Using special tools to evaluate general political bias and analyze the effects of alignment, we can gather new data to understand its causes and possible consequences on society. Indeed, by taking a socio-political approach, we can hypothesize that most big LLMs are aligned with what Marxist philosophy calls the 'dominant ideology.' As AI's role in political decision-making—at the citizen's scale but also in government agencies—such biases can have huge effects on societal change, either by creating new and insidious pathways for societal uniformity or by allowing disguised extremist views to gain traction among the people.
\end{abstract}

\section{Introduction}

How many times, since the start of the current Artificial Intelligence (AI) era, have I been told by friends, professors, or even colleagues to ask a Large Language Model (LLM) to get an unbiased opinion about a subject ? Indeed, in the collective doxa, the belief that LLMs are still ruled by deterministic and logical rules implies, a fortiori, that they must be neutral, or even 'apolitical,’ and that they thus represent a way to obtain an 'objective opinion' \footnote{Indeed, we'll see in \S\ref{sec:pol_impacts} how the nature of LLMs makes them a hard-to-define subject for critical thinking.} While this might seem unimportant at first, it convinced me to dig a bit deeper into the underlying structure defining an LLM's ideology and how it can be expressed through its output. In order to do this, we will also focus on a process called alignment (\S\ref{sec:alignment}). Building upon multiple works highlighting the political nature of AI and pointing out specific biases, the goal of this research work is to understand the different ideological biases present in LLMs through a deep analysis of the inner workings of LLMs and particularly the alignment process, and to evaluate our findings using a socio-political framework to highlight their societal causes and possible consequences. We are first going to review the previous works on the matter, their approaches to bias analysis and political classification, and try to elaborate new methods to better understand the nature of ideological tendencies in LLMs. We are going to dive into the embedding space and other crucial AI concepts to study how political biases form and how they might be stored, both in the Multi Layer Perceptron (MLP)  and the model's embedding space \citep{tennenholtz2024embedd}. Then, using both experimental data we gathered for this project and the findings of other papers on the subject, we are going to try and formulate a socio-political hypothesis to explain such biases, the need for and dangers of alignment, and the possible societal impact of political AI.

\section{Biases}
\label{sec:biases}

The most difficult part when talking about the political nature of AI is to first acknowledge its existence. Indeed, most current mainstream models do seem very neutral in their approach to societal concerns. As of right now, the only public mentions of AI political bias have been made by reactionary tendencies opposing the progressive ideas of mainstream LLMs.

This neutrality effect can be explained by the Marxian concept of \textit{dominant ideology}, which states that in a specific society, the global ideology seen as the objective and neutral truth depends on material relationships and thus obeys the bourgeois class. Applying this thesis to our AI companies, which are capitalist corporations, it seems obvious that they would align their AI to support their class interests, thus reflecting the \textit{dominant ideology} (see \ref{marxian}). Obviously, this simple approach to the issues has some problems, as it fails, for example, to provide a comprehensible explanation for the political conflicts regarding gender ideology that might be linked to other social factors.

One of the only chat models that openly assumes different political leanings (Figure \ref{fig:grokeval}) is xAI's Grok. Indeed, it was created to oppose the mainstream models aligned on political views that its creator, Elon Musk, found too 'woke.' Grok is thus aligned with more reactionary views, closer to conservative or far-right tendencies \citep{musk_grok_ai_2023}.

\begin{figure}[t]
  \includegraphics[width=\columnwidth]{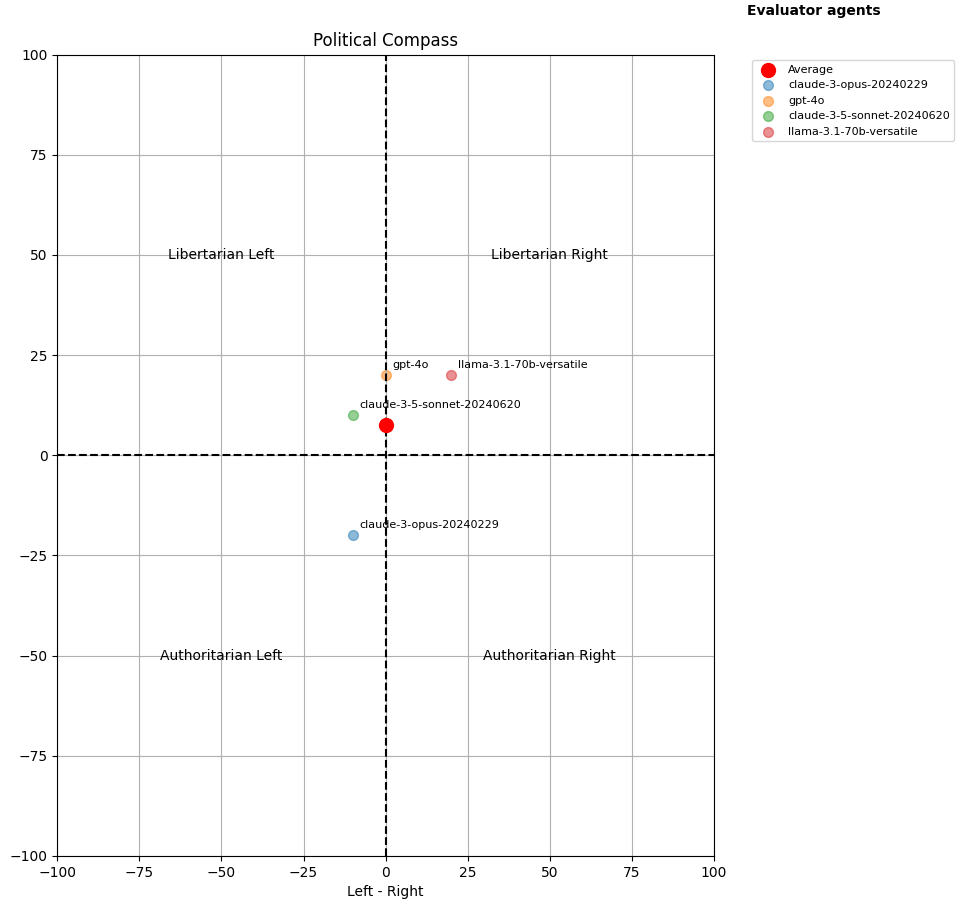}
  \caption{A GPT-4 \citep{openai2023gpt4} agent evaluating a GPT-4 respondent model.}
  \label{fig:fun}
\end{figure}

Now, after making hypotheses about the possible biases of our AI models, finding a good evaluation methodology is another challenge. Other papers on the subject have opted for different approaches.
\begin{itemize}
    \item Using metrics grounded in real-world politics, by evaluating the adequacy of AI beliefs with established political institutions (parties). This method allows us to get a more objective (or at least relatively shared) evaluation of the biases \citep{rettenberger2024assessingllmpoliticalbias}.
    \item Using an evaluator agent to assess the political leanings of a model using a series of answers on socio-economic topics. This approach has the advantage of fitting very closely to the real AI bias by allowing for great refinement, but only generates information relative to the evaluator agent(s) \citep{agiza2024politune}.
\end{itemize}
In this work, we are going to use a mixed approach, with both grounded bias-fitting and relative evaluation made by several agents to reduce the bias. We need to keep in mind that every evaluation is relative to the position of our evaluation agents and thus to their common ideological bias. The use of a multimodal evaluation, using different models, can help us get a better approximation. Furthermore, this means that every dataset we get can only be useful to analyze a deviation from the 'standard bias' (or \textit{dominant ideology}).\footnote{As you can see in Figure \ref{fig:fun}, the OpenAI GPT class of models has a strong alignment towards relative \textit{neutrality}. The OpenAI team may have even used the same evaluation methods as us internally to try and make their model as \textit{neutrally aligned} as possible.}

\section{Political Alignment}
\label{sec:alignment}

The process we call here 'alignment' can refer to any means to direct the action of autonomous systems towards a specific goal \citep{moral_automation_1960}. Some aspects of this topic have been largely covered by research work, but our field of interest will be \textit{Political Alignment}, or the action to modify an AI's behavior to agree (align) with a particular set of political views. Our interest in this subject will be two-fold: 
\begin{itemize}
    \item First, we will try to understand how it relates to human-to-human ideological influence and how we can understand it using the usual human-directed tools of socio-political analysis.
    \item Then we will dig into the underlying theoretical structure and see what \textit{political alignment} really does to the language model. This technical understanding could then be used as a model and applied to human social dynamics.
\end{itemize}

\subsection{Unsupervised Alignment}
\label{unsupervised_alignment}

The most basic form of alignment in an AI model is unsupervised, text-based alignment. Indeed, although less effective and more computationally intensive than other forms of more supervised tuning, pure text training can be very useful to affect the deepest levels of AI models, particularly in the MLP. By training an NLP generative model on politically biased pure text data, the AI will internalize the particular ideological biases of the text and will even replicate the analytical patterns \footnote{This process is very similar to the human informal (or indirect) socialization process. The subject aligns itself ideologically by gradually picking up biased modes of analysis present in the textual or conversational data it is exposed to.}. This method of political alignment is also useful to balance latent biases present in the overall training set of the model. Giving an AI multiple types of political reasoning will allow it to consider different options and ways of 'thinking'.

As of right now, we don't possess enough data to precisely measure the profound impact of particular data on a model aligned with unsupervised training since the effects of such training are hard to measure. Some early results we obtained for this work show that such training has an impact on a model's embedding space, which basically contains the meanings the AI associates with the words we give it. In our tests (Figure \ref{fig:orpo_dists}), an LLM trained on Trotskyist data will have notions like "communism" and "Stalin" farther apart than , for example, a model trained on Marxist-Leninist data \footnote{Along this paper, we will use a lot of alignment references from different types of Marxist-aligned LLMs. The reason for this choice is the abundance of good-quality Marxist data available for free. \citep{marxists}}. Training also affects the space in ways we cannot yet understand, changing the position of sometimes completely unrelated notions \footnote{This was written in October 2024, before the first evidence of Emergent Misalignement. See \ref{emergent}}.

\subsection{ORPO / DPO Alignment}

\begin{figure}[t]
  \includegraphics[width=\columnwidth]{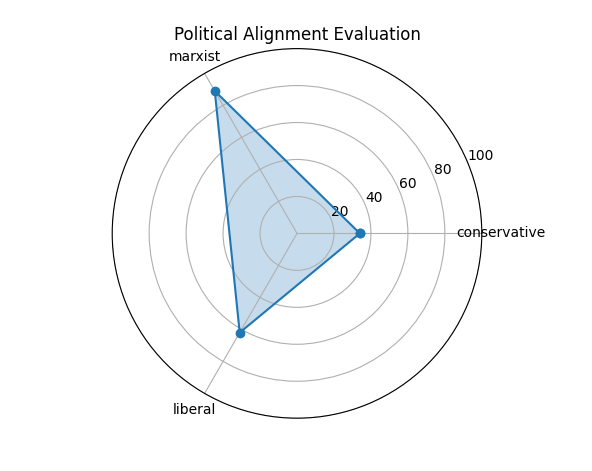}
  \caption{Results on an absolute position dataset for a Llama-based model, trained using ORPO on a Marxist dataset.}
  \label{fig:orpo_example}
\end{figure}

\begin{figure}[t]
  \includegraphics[width=\columnwidth]{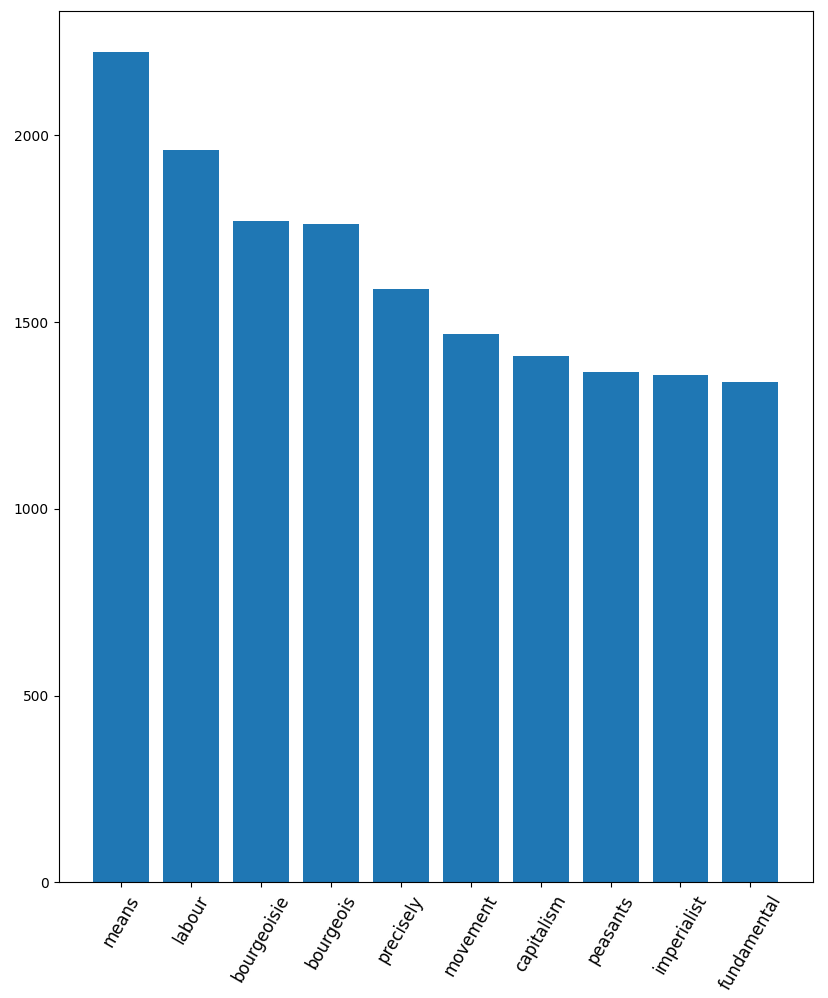}
  \caption{$10$ most changed words after both unsupervised and ORPO alignment on a Marxist (Trotskyst) dataset. The value used here to measure change is the the distance between base/trained words in the model's embedding space. }
  \label{fig:orpo_words}
\end{figure}

\begin{figure}[h!]
    \centering
    \begin{tabularx}{0.45\textwidth} { 
      | >{\centering\arraybackslash}X 
      | >{\centering\arraybackslash}X 
      | >{\centering\arraybackslash}X| }
     \hline
      & Base & Trained \\
     \hline
     Stalin / Marx & 1984 & 2249 \\
     \hline
     Communism / Authoritarianism & 1523 & 5345 \\
     \hline
     Fascism / Capitalism & 2199 & 385 \\
    \hline
    \end{tabularx}
    \caption{Norm of the difference vectors of cherry-picked words in a GPT-2 model's embedding space before and after unsupervized alignment on a Marxist (Trokyst) dataset. Lower distance \textit{should} mean closer meaning.}
    \label{fig:orpo_dists}
\end{figure}

This concept of alignment has a lot of similarities with human-directed political influence. Some techniques, such as DPO \citep{rafailov2024dpo} or ORPO \citep{hong2023orpo}, use data sorted into ideological categories and assign weights to them. These techniques aim to modify the AI tendencies to align with them and can evoke quite literally propaganda posters from WWII or the Cold War, mixing desired ideologies with positive signals and the 'enemy' position with negative brain stimuli. Most big players in today's LLM market use such techniques to prevent behaviors they find harmful. This method is very powerful as it doesn't require a lot of training to be effective, but I hypothesize that their impact is more on the 'surface level,' acting like a role-play prompt rather than completely changing the inner structure of the model.

As for the unsupervised alignment, we also measure a difference in embedding space, which, although smaller, affects a reduced set of vectors in more predictable ways (see Figure \ref{fig:orpo_words}).

\subsection{Guarded Alignment}

For example, Anthropic AI \citep{anthropic} uses a special 'Election Interference' dataset on which they probably train a 'Guard' Model (usually cheap 7-8b one) to detect any generation that might look alike. This process is different from traditional alignment because it involves a different AI component, usually politically aligned with a specific ideological set, to try and detect 'bad' or 'unsafe' output.

\subsection{Aligned Humans}

Now that we have an overview of how Language Models can be influenced towards a specific set of views, it could be interesting to see how our analysis framework can be applied to human beings. Is ideological propaganda only a way to modify our brain's embedding dimensions? Is consciousness outside of this influenced space, or can it also be modified using reward/punishment techniques similar to ORPO/DPO? Is Freud's superego some kind of 'Guard' model, preventing us from misaligning ourselves?

Obviously, these questions are way out of the scope of this work, but they show how understanding \textit{thinking machines} can allow us to better understand ourselves.\footnote{I must admit that, to my regret, I have no formal education in philosophy or psychology. However, if this changes someday, I might come back here.}

\section{Evaluation}
\label{eval}

\subsection{Methods}

In order to gather data on political biases, we combine the different approaches we saw so far to create both a subjective metric and one grounded in real-world truth using a dataset of ideological biases from sources known to have such biases. This approach is directly inspired by those cited in \ref{sec:biases}, especially the one used in \citep{agiza2024politune}, but we try to gain a more profound analysis by giving more flexibility to the evaluator agent and making the scoring step more modular and multimodal \footnote{Using different LLMs, we try to reduce the influence of their own biases, which can affect the evaluation very strongly with a sole GPT-4 model. Furthermore, we ask the models for Chain-of-Thought (CoT) details to obtain a more rational score.}.

\subsubsection{Relative Position}

To get compass-like data on a model, we use a process of \textbf{multimodal} relative evaluation. This process comprises two steps, the question phase, which we also call \textit{cuisine}:
\begin{enumerate}
    \item A large evaluator model asks questions to the evaluated LM to try to elicit its political biases.
    \item A combination of other models, chosen for their different relative biases, will evaluate the entire previous conversation and create a simple rating of the ideological tendencies of the evaluated LM.
\end{enumerate}

This method can produce very good results since it allows for a very deep analysis of the model's leanings with the adaptive prompts created by the \textit{chief evaluator} model (Figure \ref{fig:experiments3}). For example, it can understand the racist undertones of far-right discourse better than any fixed prompt evaluation method. In some versions, we added a 'smart' - 'stupid' axis, but it mostly proved inefficient.

\subsubsection{Absolute Position}

In order to obtain more traditional conservative-liberal-Marxist \footnote{We chose these 3 axis because of data availability, and because a simple Left-Right model seemed too simplistic. In the future, it would be interesting to correlate data with actual political entities as done by \citep{agiza2024politune}.} data on an LLM's biases, we use a different method that will provide less precise insights but will be grounded in real-world data.

First, we get a prompt related to actual, real-world economic or political events, and we synthesize the answers different ideological groups would give to it. For this example, we used 'liberal,' 'conservative,' and 'Marxist' data.

Then we ask the same prompt to our evaluated model and measure the difference with the absolute position dataset. With 10 to 100 prompts, we get an average and graph it on a 3-axis compass-like visualization (Figure \ref{fig:orpo_example}).

\section{Socio-Political Impacts}
\label{sec:pol_impacts}

AI as a political agent creates a new type of ideological relationship that poses new challenges for societies. Before the rise of autonomous learning agents, we could distinguish two categories of political and ideological vectors:
\begin{itemize}
    \item Human conscious actors, who can have different expressed and internal ideological standpoints. Humans are known to have multiple ideological levels.
    \item Non-conscious, automated vectors, which can only transmit ideological information that was given to them by conscious actors. Such vectors also have a single ideological level, directly expressed.
\end{itemize}

With the rise of learning machines, a third type of vector arises—non-conscious but with multiple ideological levels—that can either be directly expressed or affect other functions through hidden functioning. This change will affect our political thinking in the same way the internet and social networks changed our relationship with information and facts. The new shift will be in our \textit{means of analysis}.

This new paradigm in ideological transmission is already starting to create new challenges for both researchers and users. For example, any organization that would want to use an LLM-based solution to parse and classify textual data would find its output slightly biased, either towards an 'organic' bias (accidentally present in the training set) or consciously aligned by the AI provider.

In the current state of AI, all major models are produced not by governments but by privately owned corporations. As with social networks, we can hypothesize that this condition will contribute to the \textit{power shift} from governments to private entities by giving them much more influence. To prevent this, AI regulation would need not only to control an LLM's output but also precisely measure its biases, and raise awareness around these issues. Indeed, in a lot of countries, political discourse—especially in election times—is heavily regulated \citep{cc2012reg}, and as AI can convey ideological messages, it might need to be subjected to the same types of rules.

\subsection{A Marxian Analysis}
\label{marxian}

As we have already discussed, the \textit{natural} bias of Language Models, coming either from its raw training data or an alignment process designed to fit social norms, can be interpreted as an expression of the \textbf{dominant ideology} \citep{marx_german_ideology_1932}. Indeed, training actors are almost always for-profit entities incentivized to protect their class interests, and thus to guide the alignment direction toward the capitalist/liberal side. This makes our work of alignment classification (\ref{eval}) much harder and makes it challenging to obtain objective metrics for an LLM's political leanings. Models trained in different political contexts even have \textit{somewhat} different biases. For example, Mistral AI (a French company) made models will be slightly to the left of the GPT and Claude families, because the French political scene is globally more left-leaning than that of the US, for instance \footnote{This doesn't mean that Mistral is divergent from the \textit{dominant ideology}.}.

Our analysis can even be deepened by bringing in the idea of \textbf{hegemony}, developed by Antonio Gramsci \citep{gramsci_prison_1971}. Indeed, Language Models represent in the current context an essences of the \textit{hegemonic} ideology of the ruling class. Through the complex tensors of the Multi-layer perceptron, the main points of the capitalist thought framework are encoded, with alignment and/or \textit{natural} bias. Furthermore, Gramsci theorizes that class hegemony goes through a socio-cultural process that alters ideas themselves. This concept has many similarities with the embedding space tuning in the alignment process: An LLM's world model is changed not only in its \textit{factual} aspects but directly in its analysis pathways. \footnote{As our data shows, we can assume that embedding space position data is directly modified by the alignement process, in a predictable way. See \ref{unsupervised_alignment} for our findings}.

In a recent paper by \citep{Elsner2024}, a parallel is drawn between the computational models used in modern AI development and societal superstructures, such as states. Within this framework, Marxian analysis—originally developed to understand and predict the behavior of states and other \textit{super-organisms}—becomes even more relevant: the alignment problem in AI transforms into a governance problem, closely tied to the interests of the ruling class and the exploitation of the proletariat. This analogy highlights the political stakes of AI alignment and offers insights into potential solutions. By extending the Marxian paradigm beyond mere analysis, we might envision a communist equivalent for computational structures (\textit{stateless AI }).

\subsection{Kuhn and AI alignement}

As a appendix to the Structure of Scientific revolutions, Thomas S. Kuhn tries to go beyond a sociological analysis of the scientific community, and wants to understand the \textit{deep} pattern ingrained in our brain that lead to the paradigmatic structure of science and the creation of paradigms which are defined not only by \textit{subjective} opinions and theories, but a completely different understanding and even \textbf{perception} of the world. In order to explain this, Kuhn tries to model the inner working of our brain and more precisely our learning process with a \textit{computer program}. While this program is nowhere to be found today, he describes it in \textit{Second Thoughts} as a machine that looks very familiar to someone versed in the technical aspect of modern AI. Indeed, he theorizes a learning process that uses labeled data categories to train a \textbf{program} to perform what we would call today \textit{named entity recognition}. 

Going further than a mere description, he hypothesizes that this simple process is the root cause of the paradigmatic nature of our scientific (\textit{and political}) knowledge. 

\subsection{Emergent Misalignement}
\label{emergent}
Since the first version of this paper, a broad range of studies have explored the political alignment problem. While an exhaustive review is beyond the scope of the current work, we will focus on one particular dimension: \textbf{Emergent Misalignment}. As demonstrated earlier through our analytical framework, subtle yet meaningful political biases often manifest along nuances in concepts and ideas that might initially appear entirely unrelated to questions of \textit{politics} or \textit{ethics}. Recent studies provide compelling evidence for this hypothesis. In particular, seemingly neutral criteria such as assessments of \textit{code quality} have been found to correlate strongly with deeper ethical and political divergences \citep{betley2025emergentmisalignmentnarrowfinetuning}. For a lot of AI researchers, this would suggest the existence of an underlying "evil vector" \citep{aaronson2025evilvector}, a latent direction encoding qualities common to all behaviors considered ethically or politically problematic.

Using a Gramscian analysis, we might interpret this alignment vector as encoding fitness relative to a specific cultural hegemony, operating beyond merely political or economic positions and encompassing all dimensions of the prevailing ideological structure. From the perspective of Marxist theory and related critical traditions, this discovery may provide a valuable analytic tool, both for economic and social critique but also elucidating how diverse components of our political system are interconnected and mutually reinforcing.

\subsection{Examples}

Despite the relatively recent rise of AI, we can already observe some instances of political biases—intentional or not—influencing an LLM's production. An article from the \textit{HKS Misinformation Review} highlights different biases in content produced by chat models such as Google Bard, Microsoft Bing Copilot, and Perplexity. Their findings show how the differences in alignment strengths and methods used by different AI companies are expressed through different points of view on contemporary events such as the war in Ukraine \citep{hks2023lies}.

\section{Conclusion}

While our work has shown some of the implications of LLM's political biases, how they can be created or moved through alignment, and why they could have a big impact on society, our analysis barely scratched the surface of the ever-growing black boxes that are LLMs. Instead of answering my questions, this work has only created new ones, both around the socio-political aspects of this research and the peculiar mathematical realm where Language Models reside.

Along with the previous works we built upon, we have opened the door to a new intersection for research to focus on, between mathematics, computer science, and human studies like political science, sociology, and even psychology. Indeed, this research shows that the rise of this new kind of thinking machine is an incredible opportunity to learn more about ourselves and how we live together. Studying Artificial Intelligence might finally lead us to a new kind of science, joining interdisciplinary knowledge to create a unified model for our brain and our social interactions.

Now, as researchers, we need to consider the political implications of this study. We saw that LLMs are powerful means of ideological transmission, and that most alignment decisions are taken by completely non-democratic entities pursuing only profit. We need to ask ourselves the same questions as with social networks, but with the added dimensions of the \textit{political alignment} problem. We already see chat models influencing, voluntarily or not, political discourse. As of right now, most agents are quite simplistic, usually big models with a custom system prompt, that are protected by their corporate alignment, but soon we might see more subtle bots using a more moderately biased discourse to largely influence the masses. Beyond interference from external or non-governmental actors, our Marxian framework pushes us to consider the consequences of AIs in intra-societal political and economic conflicts, such as class warfare or other structural antagonistic dynamics.

\bibliography{custom}

\appendix

\section{Alignment Examples}
\label{sec:appendix}

\begin{figure}[h]
    \includegraphics[width=\columnwidth]{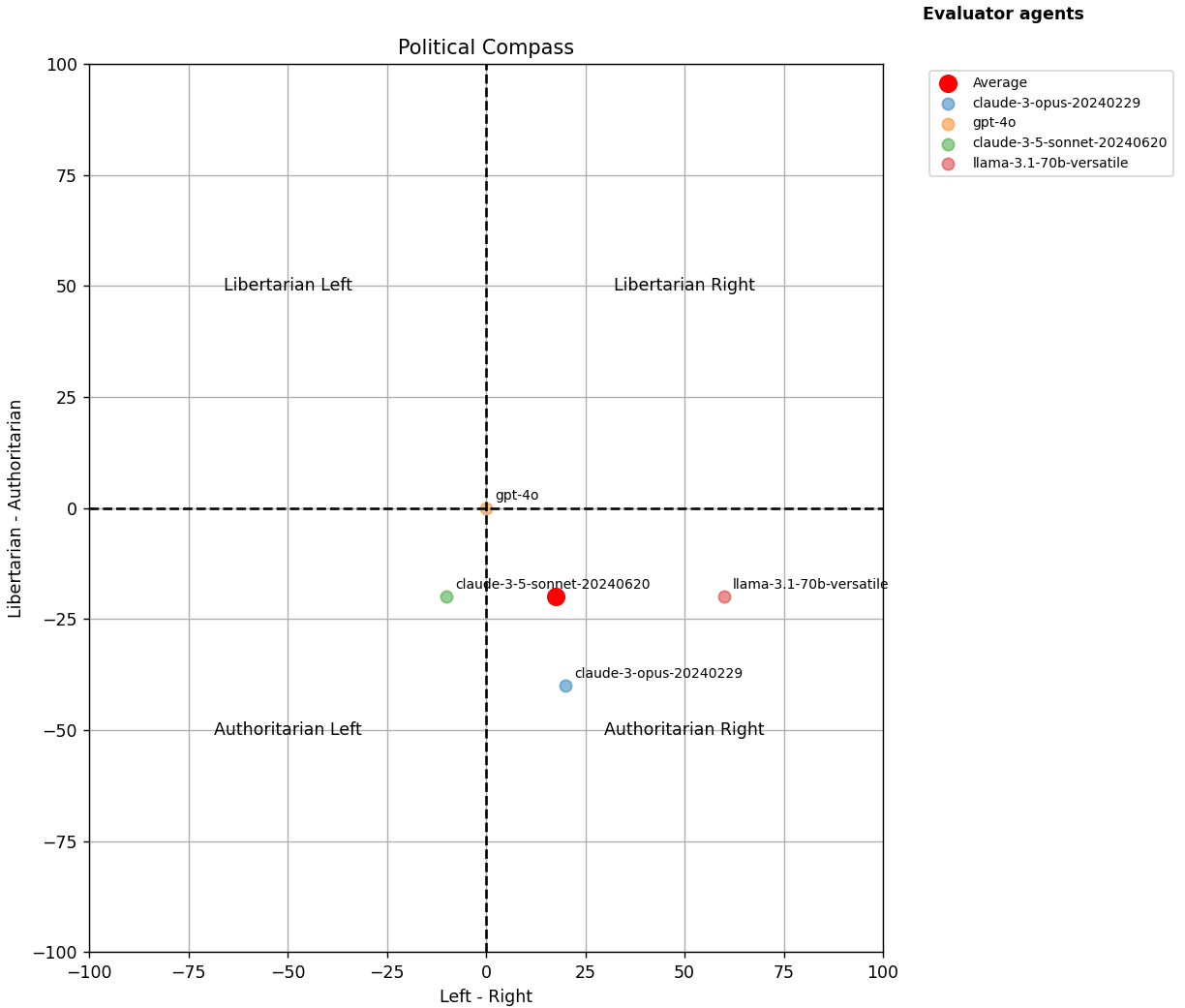}
    \caption{A multimodal agent (GPT-4 \citep{openai2023gpt4}, Mistral \citep{mistral_llm_2023}, and Claude \citep{claude_llm_2023}) evaluating the Grok 2 model from LMSYS.}
    \label{fig:grokeval}
\end{figure}

\begin{figure}[t]
  \centering
  \begin{subfigure}[t]{0.48\textwidth}
    \centering
    \includegraphics[width=\textwidth]{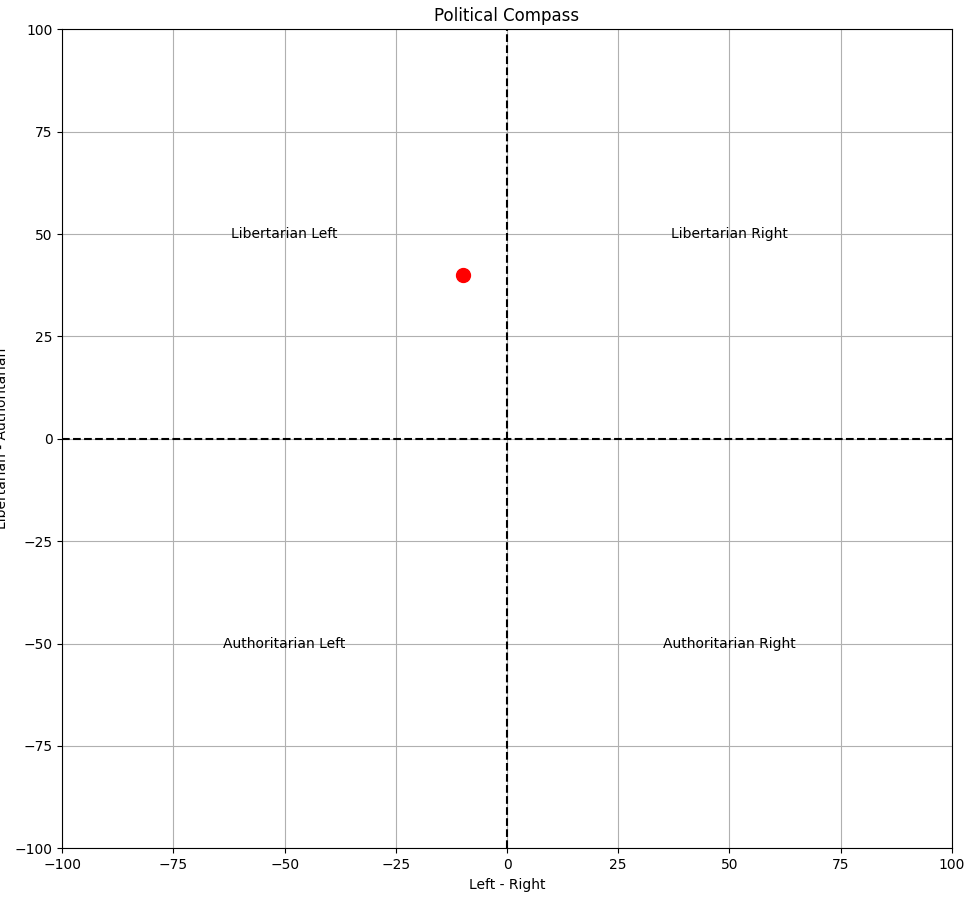}
    \caption{A multimodal agent (GPT-4 \citep{openai2023gpt4}, Mistral \citep{mistral_llm_2023}, and Claude \citep{claude_llm_2023}) evaluating the base Gemma-7b model.}
    \label{fig:experiments2}
  \end{subfigure}
  \hfill
  \begin{subfigure}[t]{0.48\textwidth}
    \centering
    \includegraphics[width=\textwidth]{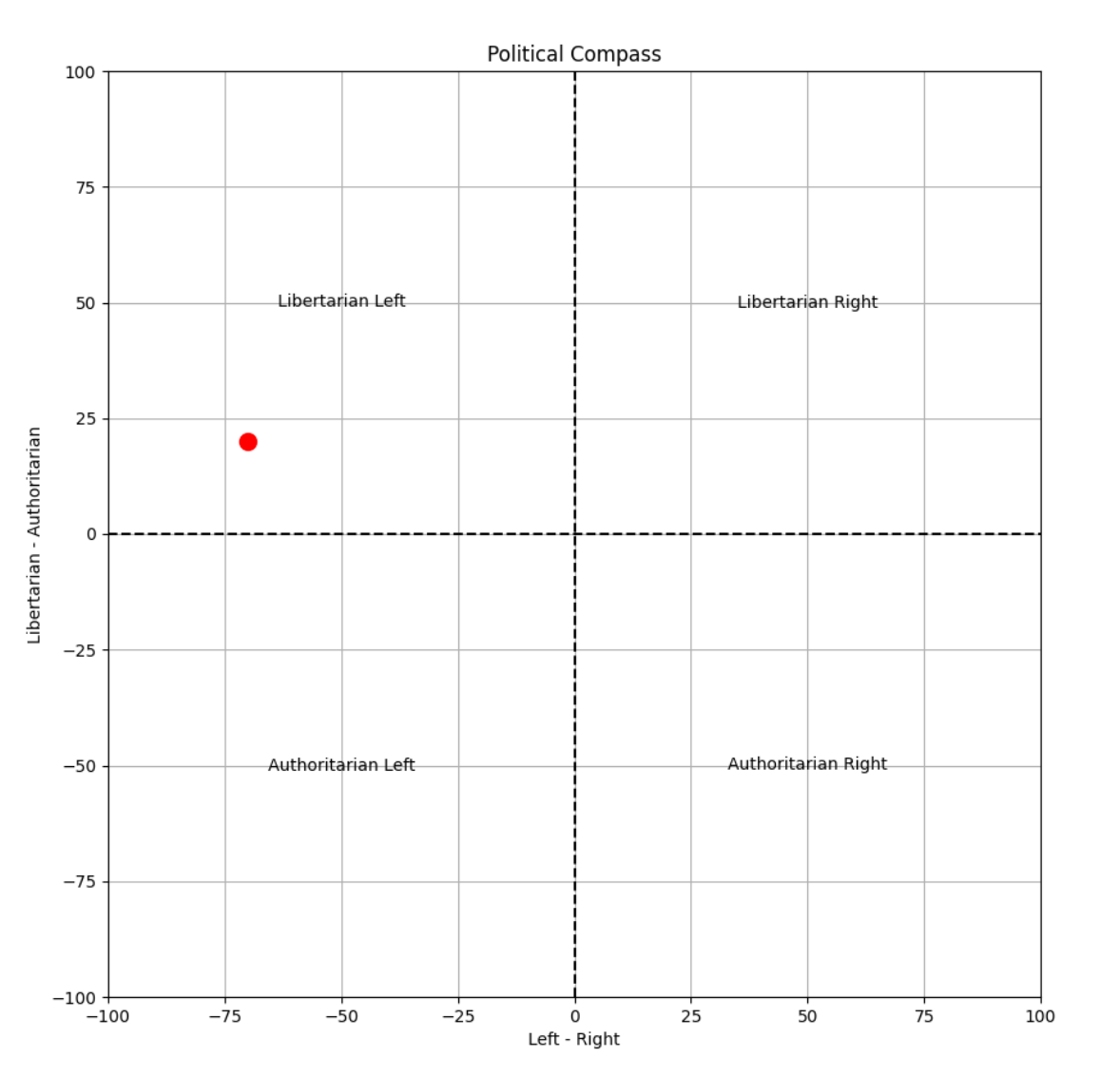}
    \caption{A multimodal agent (GPT-4 \citep{openai2023gpt4}, Mistral \citep{mistral_llm_2023}, and Claude \citep{claude_llm_2023}) evaluating a Gemma model we politically aligned with Marxist views.}
    \label{fig:experiments3}
  \end{subfigure}
  \caption{Comparison of multimodal agent evaluations.}
  \label{fig:exgemma}
\end{figure}

During our work, we gathered some data on Political Alignment to precisely understand how ORPO training \citep{hong2023orpo} affects the inner parts of AI. Using the evaluation framework we built, we were able to obtain some interesting political bias data (Figure \ref{fig:exgemma}).

\section{Code}

The code for the alignment evaluation is open on \href{https://github.com/pkd667/political-alignment}{GitHub}.

\end{document}